\documentclass[12pt,fleqn]{article}
\usepackage{amssymb}
\usepackage{amsmath}
\usepackage{amstext}

\usepackage{times}
\usepackage{url}
\usepackage{amsfonts,amsmath,amssymb,amsthm}
\usepackage[authoryear,semicolon,longnamesfirst,sectionbib]{natbib} 

\usepackage{lscape}
\usepackage{rotating}

\usepackage{float,graphicx, amssymb, amsthm, amsmath, amsfonts, verbatim, latexsym}
\usepackage{setspace}
\usepackage[caption=false,labelformat=simple,hangindent=0pt,format=hang,singlelinecheck=false,position=top,topadjust=-130pt]{subfig}

\newcommand{\overallsubfigwidth}{.335\textwidth}

\usepackage{cos_macros}

\usepackage{multirow}
\usepackage{array}
\usepackage[caption=false]{subfig}
\usepackage{color}

\begin{document} 

\title{Bias and high-dimensional adjustment in observational studies of peer effects$^*$}

\author
{Dean Eckles$^{1}$ and Eytan Bakshy$^{2}$
\\
\normalsize{$^{1}$Massachusetts Institute of Technology, $^{2}$Facebook}\\
}

\date{}

\maketitle 

\begin{abstract}
Peer effects, in which the behavior of an individual is affected by the behavior of their peers, are posited by multiple theories in the social sciences. Other processes can also produce behaviors that are correlated in networks and groups, thereby generating debate about the credibility of observational (i.e. nonexperimental) studies of peer effects. Randomized field experiments that identify peer effects, however, are often expensive or infeasible. Thus, many studies of peer effects use observational data, and prior evaluations of causal inference methods for adjusting observational data to estimate peer effects have lacked an experimental ``gold standard'' for comparison. Here we show, in the context of information and media diffusion on Facebook, that high-dimensional adjustment of a nonexperimental control group (677 million observations) using propensity score models produces estimates of peer effects statistically indistinguishable from those from using a large randomized experiment (220 million observations). Naive observational estimators overstate peer effects by 320\% and commonly used variables (e.g., demographics) offer little bias reduction, but adjusting for a measure of prior behaviors closely related to the focal behavior reduces bias by 91\%. High-dimensional models adjusting for over 3,700 past behaviors provide additional bias reduction, such that the full model reduces bias by over 97\%. This experimental evaluation demonstrates that detailed records of individuals' past behavior can improve studies of social influence, information diffusion, and imitation; these results are encouraging for the credibility of some studies but also cautionary for studies of rare or new behaviors. More generally, these results show how large, high-dimensional data sets and statistical learning techniques can be used to improve causal inference in the behavioral sciences.

* We are grateful to L. Adamic, S. Aral, J. Bailenson, J. H. Fowler, W. H. Hobbs, G. W. Imbens, S. Messing, C. Nass, M. Nowak, A. B. Owen, A. Peysakhovich, B. Reeves, D. Rogosa, J. Sekhon, A. C. Thomas, J. Ugander, and participants in seminars at New York University Stern School of Business, Stanford University Graduate School of Business, UC Berkeley Department of Biostatistics, Johns Hopkins University Bloomberg School of Public Health, University of Chicago Booth School of Business, Columbia University Department of Statistics, and UC Davis Department of Statistics for comments on this work. D.E. was previously an employee of Facebook while contributing to this research and is a contractor with Facebook. D.E. and E.B. have significant financial interests in Facebook.
\end{abstract}
\pagebreak

\section{Introduction}
Understanding how the behavior of individuals is affected by the behavior of their peers is of central importance for the social and behavioral sciences, and many theories suggest that positive peer effects are ubiquitous \citep{sherif_psychology_1936,blume_statistical_1995,centola_complex_2007,granovetter_threshold_1978,manski_economic_2000,montanari_spread_2010}.
However, it has been difficult to identify and estimate peer effects \emph{in situ}. Much of the most credible evidence about peer effects in humans and primates comes from small experiments in artificial social environments \citep{asch_studies_1956,sherif_psychology_1936,whiten_conformity_2005,herbst_peer_2015}.
In some cases, field experiments modulating tie formation and group membership \citep{sacerdote_peer_2001,zimmerman_peer_2003,lyle_estimating_2007,carrell_does_2009,centola_spread_2010,firth_pathways_2016}, 
shocks to group or peer behavior \citep{aplin_experimentally_2015, banerjee_diffusion_2013, bond_massive_2012, cai_social_2015, eckles_estimating_2016, van_potent_2013}, or subsequent exposure to peer behaviors \citep{aral_creating_2011,bakshy_social_2012,salganik_experimental_2006}
have been possible, but in many cases these experimental designs are infeasible.
Thus, much recent work on peer effects uses observational data from new large-scale measurement of behavior \citep{aral_distinguishing_2009,bakshy_everyones_2011,friggeri_rumor_2014,ugander_structural_2012,allen_network_2013} or longitudinal surveys \citep{christakis_spread_2007, iyengar_opinion_2011, banerjee_diffusion_2013, card_peer_2013, christakis_social_2013, fortin_peer_2015}.
Many of these studies are expected to suffer from substantial confounding of peer effects with other processes that also produce clustering of behavior in social networks, such as homophily \citep{mcpherson_birds_2001} and external causes common to network neighbors. Thus, even when issues of simultaneity \citep[including ``reflection'',][]{manski_reflection_1993} can be avoided, it is thus generally not possible to identify peer effects using observational data without the often suspect assumption that adjusting for available covariates is sufficient to make peer behavior unconfounded \citep[i.e., as-if randomly assigned;][]{shalizi_homophily_2011, angrist_perils_2014}.
However, even if these assumptions are not strictly satisfied, some observational estimators, especially those that adjust for numerous or particularly relevant behavioral variables, may have relatively small bias in practice, such that the bias is small compared with other sources of error (e.g., sampling error) or is small enough to not change choices of theories or policies.

Using a massive field experiment as a ``gold standard'', we conduct a \emph{constructed observational study} by adding a nonexperimental control group to a randomized experiment \citep{lalonde_evaluating_1986,dehejia_causal_1999,dehejia_propensity_2002,hill_comparison_2004} to assess bias in observational estimators of peer effects in the diffusion of information and media, which has been widely studied \citep{katz_personal_1955,wu_who_2011,berger_arousal_2011,myers_information_2012, bakshy_role_2011,bakshy_exposure_2015,friggeri_rumor_2014,cheng_can_2014,flaxman_filter_2016,goel_structural_2015}. Diffusion of information and media, especially via Internet services, is now central to multiple topics in applied research, including studies of product adoption \citep{goel_structural_2015} and political participation \citep{friggeri_rumor_2014, bakshy_exposure_2015,barbera_tweeting_2015,flaxman_filter_2016,allcott_social_2017}.
The present work is the first to experimentally evaluate state-of-the-art estimators for observational studies of peer effects, and it does so in a setting of substantial and increasing relevance.

We review related work in the next subsection. Section \ref{sec:data} describes the randomized experiment, observational data, and estimators used in this study. Section \ref{sec:results} compares the resulting experimental and observational estimates of peer effects, where we find substantial variation among the performance of the observational estimators.
\subsection{Related work}
Prior evaluations of observational estimators of peer effects lacked comparison with an experiment and instead relied on sensitivity analysis \citep{vanderweele_sensitivity_2011}, simulations \citep{thomas_social_2013}, and analyses when the absence of peer effects is assumed \citep{cohen-cole_detecting_2008}. For this reason, we review prior work on peer effects in this section, but also consider observational--experimental comparisons in other areas of study. While causal inference for peer effects faces distinctive threats to validity (e.g., homophily), the present study also has methodological advantages over many prior observational--experimental comparisons for educational, medical, and public policy interventions \citep[e.g.,][]{lalonde_evaluating_1986,heckman_matching_1997,dehejia_causal_1999,dehejia_propensity_2002, hill_comparison_2004,michalopoulos_can_2004, diaz_assessment_2006,shadish_can_2008}.
Unlike other constructed observational studies (where, e.g., different survey questions were used for the experimental and nonexperimental data \citep{diaz_assessment_2006}), here measures and outcomes for the experimental and nonexperimental data are identically defined and measured. Furthermore, unlike prior ``double-randomized'' designs \citep{shadish_can_2008}, this study has sufficient statistical power to detect confounding bias and examines processes that cause exposure \emph{in situ}. Here we are able to study peer effects across many distinct behaviors and evaluate high-dimensional adjustment for covariates that can be constructed from routinely collected behavioral data, rather than custom-made survey instruments \citep{diaz_assessment_2006,shadish_can_2008,pohl_unbiased_2009}. In the remainder of this section, we make more detailed comparisons with such prior evaluation of observational methods outside the study of peer effects. We focus on single-study comparisons, where the prior work is most similar to the present contribution. We highlight some of the sources of ambiguity in this prior work and note the advances of the present study with respect to methodology and applicability to contemporary research.
\subsubsection{Simulations}
Simulations can be used to evaluate observational methods under known models for selection into the treatment. Of course, simulations require assuming both selection and outcome models, which are typically unknown for the circumstances of interest. Simulations have been used to evaluate the consequences of adjustment for additional covariates \citep{steiner_bias_2015}, which need not always be bias reducing \citep{ding_adjust_2015}. In the context of peer effects, simulations have been used to illustrate how peer effects are confounded with homophily  \citep{shalizi_homophily_2011} and the consequences of specific modeling choices  \citep{thomas_social_2013}. Realism can be somewhat improved by using real data as a starting point for a simulation, e.g., by only simulating the missing potential outcomes  \citep[e.g.,][]{imai_causal_2004} but using real covariates and treatment data. Nonetheless, simulations are limited in their ability to tell us about the magnitude of bias and bias reduction in practice.
\subsubsection{Meta-analyses}
Meta-analyses of observational and experimental studies are sometimes used to assess the bias of observational methods. For example,  \citet{hemkens_agreement_2016} compare experimental and observational studies of effects of several clinical treatments on mortality. Similarly,  \citet{schuemie_interpreting_2014} examine numerous studies using routinely collected data and demonstrates how dependent observational estimates can be on specific modeling choices. Meta-analyses often face the problem that the observational and experimental studies differ in numerous ways besides how units were assigned to treatment.
These differences include the populations sampled from, the implementation of the treatment, and the outcomes measures may all differ.
In the context of peer effects, meta-analysis is especially difficult because of the small number of field experiments previously conducted. One recent meta-analytic study of peer effects in worker productivity  \citep{herbst_peer_2015} is able to compare lab experiments with field studies, but the field studies are largely not randomized experiments; they do not attempt a comparison of field experiments and observational studies.\footnote{
There are two field studies with random assignment. Both use random assignment of peers, rather than random assignment of exposure to peer behavior.  \citet{goto_incentives_2011} estimate peer effects in the productivity of rice planters; this estimate is the largest of any lab or field study in the meta-analysis.  \citet{guryan_peer_2009} estimate peer effects in the performance of professional golfers; they find no significant peer effects. These distinctive settings and results illustrate the challenge of conducting a meta-analysis such that randomized and observational studies are comparable.}  To our knowledge, the experiment we use is the only large field experiment identifying peer effects in online information diffusion. So a meta-analysis here would be limited to comparing a single experimental estimate (or a small number of estimates) to a number of observational estimates. 
\subsubsection{Single-study comparisons}
Single-study comparisons --- such as the present work --- involve comparing the results from a single study that includes observations where some units are randomly assigned and other units are not randomly assigned. Among single-study comparisons, one can distinguish between studies that combine such data by different means.

{ \bf Constructed observational studies.}
First, there are studies in which an existing randomized experiment is augmented by adding observational data --- usually by adding a non-experimental control group (NECG); these are sometimes called \emph{constructed observational studies}  \citep{hill_comparison_2004,hill_comment_2008}, since an observational study is constructed by the addition of data to an experiment. Prominent examples of constructed observational studies have compared observational and experimental estimates of effects of job training programs  \citep{lalonde_evaluating_1986,heckman_matching_1997,dehejia_causal_1999,dehejia_propensity_2002}, social welfare interventions  \citep[e.g., conditional cash transfers, ][]{diaz_assessment_2006}, welfare-to-work programs \citep{michalopoulos_can_2004}, and medical interventions  \citep{hill_comparison_2004}). There are a small number of papers that conduct quantitative  \citep{heinsman_assignment_1996,glazerman_nonexperimental_2003} or qualitative  \citep{michalopoulos_can_2004,cook_three_2008,wong_empirical_2016} reviews of these within-study comparisons, though the set of available studies to review has been dominated by studies of job-training and other educational treatments.

The results of these constructed observational studies have often been ambiguous for multiple reasons. First, different choices of models and theory-driven criteria for excluding units yield dramatically different comparisons with the experimental estimates. For example, in evaluating a job training program,  \citet{dehejia_propensity_2002} exclude units that do not have earning data for two prior periods, based on theoretical reasons to expect a dip in earnings right before the decision to enroll; other authors have highlighted how consequential this decision is for the results  \citep{smith_does_2005,dehejia_practical_2005,hill_comment_2008}.
Second, while it has been common to interpret observational--experimental discrepancies as due to bias in the observational estimators, there are often other explanations.  Though constructed observational studies aim to avoid the incomparability of observational and experimental estimates that occurs in meta-analyses, \citet{shadish_can_2008} argue that many constructed observational studies ``confound assignment method with other study features'' (p. 1335), such as the places or times data is sampled from, the implementation of the treatments, the version of the measures used as covariates or outcomes, and different patterns of missing data. For example, while  \citet{diaz_assessment_2006} find experimental--observational discrepancies in estimated effects of a conditional cash transfer program, for many of the outcomes these can be attributed to differences in the survey measures used for the experimental and observational data. Despite these limitations, of the six observational--experimental comparisons they review,  \citet{cook_three_2008} categorize  \citet{diaz_assessment_2006}, along with  \citet{shadish_can_2008}, as one of the two less ambiguous comparisons.

{\bf Doubly randomized preference trials.}
Second, there are studies in which units are randomly assigned to whether they will be randomly assigned to treatment or whether another process (e.g., self-selection when given a choice) will determine their treatment. This design is sometimes called a \emph{doubly randomized preference trial} (DRPT)  \citep{long_causal_2008,shadish_can_2008}.
 \citet{shadish_can_2008} motivate its use for evaluating observational methods by noting some of the shortcomings of constructed observational studies we have described above: a DRPT involves only varying the assignment method, while holding constant the population, treatment implementation, and measures. 

Papers reporting on DRPTs have argued they provide evidence about the bias of observational estimators and which types of covariates and analysis methods most reduce that bias.  \citet[][p. 1341]{shadish_can_2008} conclude that ``this study suggests that adjusted results from nonrandomized experiments can approximate results from randomized experiments.''  \citet[][p. 260]{steiner_importance_2010} describe their analysis as having ``identified the specific contributions'' made by different sets of covariates.  \citet{pohl_unbiased_2009}, who conduct a smaller replication of this DRPT, conclude that ``we have shown that it is possible to model the selection process and to get an unbiased treatment effect even with nonrandomized experiments'' (p. 475) and that ``the
choice of covariates made all the difference, and the mode of data analysis [regression adjustment, propensity score methods] did not'' (p. 474).

DRPTs are a useful tool for evaluating observational methods, but the DPRTs conducted to date have important threats to their statistical and external validity. The two DRPTs described above have relatively small samples (n = 445 and n = 202) for making a comparison between estimates computed on disjoint halves of the data. The conclusions reviewed above are apparently not based on formal statistical inference: for  \citet{shadish_can_2008} and  \citet{steiner_importance_2010} the experimental and unadjusted observational estimates are statistically indistinguishable, so their results are ``basically descriptive''  \citep[p. 256]{steiner_importance_2010}. The same is true of the smaller replication by  \citet{pohl_unbiased_2009}.
Thus, there is no evidence of any bias to be reduced or eliminated. Of course, depending on how much bias was expected, this could be notable on its own. While these papers do not report such tests, the comparisons \emph{among} different observational estimators will generally have greater power. A reanalysis of the results reported in  \citet{steiner_importance_2010} suggests that at least some of the observational estimators are converging to different estimands; see Supplementary Information Section 5.
Thus, if these studies do provide formal statistical evidence for bias and bias reduction, it is through comparison of multiple \emph{observational} estimators that would all converge to the same quantity in the absence of bias; that is, the randomized experiment arm of these particular DRPTs does not contribute evidence of bias or bias reduction.

Besides issues of statistical validity, DRPTs can lack external validity and relevance to use of routinely collected data for social science. In addition to using convenience samples of college students, many of the factors that are expected to contribute to confounding bias are absent from the self-selection process in the non-randomized arms of the DRPTs. That is, the participants are presented with a choice between two alternative activities (math and vocabulary exercises), so it is not that some people are unaware of some of the choices, unable to choose them because of costs, etc. This simplification is especially notable for rare treatments. In the context of peer effects, where the treatment is exposure to, e.g., a friend's adoption of a product or sharing of a URL, such rarity is common.

{\bf Differences with the present study.}
We have described the present study as a constructed observational study, in that we graft observational data (the NECG) onto a randomized experiment. Like some other constructed observational studies, the present study has the advantage of evaluating observation methods when applied to the actual processes of selection into treatment (i.e., exposure to a peer behavior) \emph{in situ}.

The present study also enjoys some of the advantages of DRPTs. Like DRPTs, the experimental control group and the non-experimental control group are drawn from the same population and all of the variables are measured in the same way. In fact, our NECG consists of individuals and items drawn from the same population of individuals and items in the experiment.
Thus, the main explanation of any observational--experimental discrepancies is a process that causes users to be selected into exposure to URLs they are more (or less) likely to share --- that is, confounding by, e.g., homophily, common external causes, and past influence.

Because there are millions of distinct URLs that can be grouped into thousands of domains, each of which is analyzed separately, the present study also allows for a form of internal replication that makes it similar to meta-analysis. Other single-study comparisons have usually only studied a single contrast between two treatments.

The present study is of particular relevance to other contemporary empirical research that makes use of new sources of routinely collected data about human behavior, such as from Internet services and economic activity. This is in contrast to some of the prior work that made use of custom purpose-built measures (e.g., measures of preferences for one treatment over another, as in DRPTs) that are of less relevance to use of routinely collected data.
\section{Data and method}
\label{sec:data}
We analyze a large experiment that randomly modulated the primary mechanism of peer effects in information and media sharing behaviors on Facebook: the Facebook News Feed \citep{bakshy_role_2011}.
Facebook users can share links to particular Web pages (URLs), which is a common way of disseminating news, entertainment, and other information and media.
Using a cryptographic hash function, a small percentage of user--URL pairs were randomly assigned to a \emph{no feed} condition in which News Feed stories about a peer sharing that URL were not displayed to that focal user. Deliveries of these stories and held out deliveries (i.e., those for pairs in the \emph{no feed} condition) were recorded. 
Taking exposure to a peer sharing a URL as the treatment, this experiment identifies peer effects for users who would have been exposed to peer sharing. These peer effects may arise from multiple processes, including peer sharing exposing egos to novel information, as is typical in many diffusion and product-adoption settings. Unlike other work \citep{bakshy_social_2012} that holds the content fixed and only considers the presence of a peer's name being linked to the content, this experiment targets an estimand that combines these effects.
Key causal quantities identified by the experiment include the relative risk of sharing,
$
RR = p^{(1)} / \; p^{(0)},
$
and the risk difference of sharing (i.e., the average treatment effect on the treated),
$
\delta = p^{(1)} - \; p^{(0)},
å$
where $p^{(1)}$ is the probability of sharing a particular URL when exposed to a peer sharing that URL for those that would be exposed and $p^{(0)}$ is the probability of sharing a particular URL when not exposed to a peer sharing that URL for those that would be exposed. Note that $p^{(0)}$, but not $p^{(1)}$, involves a counterfactual.

\subsection{Propensity score modeling and stratification}
All of the observational estimators evaluated here are the result of post-stratification (i.e., subclassfication) by the domain name of the URL (e.g., for the URL \url{http://www.cnn.com/article_x}, the domain is \url{www.cnn.com}); that is, per-domain estimates are combined weighting by the number of exposed observations per domain. 
The adjusted estimators additionally use granular stratification on estimated propensity scores \citep{rosenbaum_central_1983,rosenbaum_reducing_1984,rubin_estimating_1997}. 
The propensity score $e(X_{iu}) = \Pr(E_{iu} = 1 \vert X_{iu})$ is the probability that user $i$ is exposed to a peer sharing URL $u$, where $X_{iu}$ are variables describing that user--URL pair \citep{rosenbaum_reducing_1984}. In an observational study, researchers typically rely on the following assumptions. Under \emph{conditional unconfoundedness}, the potential outcomes are independent of the exposure, $(Y_{iu}(0), Y_{iu}(1)) \perp | X_{iu}$. Under \emph{overlap} or \emph{positivity}, units have positive conditional probability of exposure and non-exposure, $e(X_{iu}) \in (0, 1)$.
Conditional unconfoundedness implies that exposure is also unconfounded conditional on $e(X_{iu})$.
In observational studies, propensity scores are estimated using available covariates and conditioned on using regression adjustment, matching, weighting, or post-stratification.
We estimate propensity scores using $L_2$-penalized logistic regression (i.e., logistic ridge regression) using different sets of predictors. We set the $L_2$ penalty $\lambda = 0.5$. The effect of the penalty on the resulting estimates is expected to be small for two reasons. First, the estimated scores are used for stratification, so only the rank of the scores matters for the analysis;
thus, the size of the penalty primarily serves to control how much more small principal components are shrunk than the larger principal components. 
Note that for linear ridge regression with a univariate or orthonormal basis, the penalty has no effect on the ranks of the scores.
Second, most of the models have many more observations than input dimensions; even for models with $\mn{M}$, since this matrix is sparse, for domains with a small number of observations, only some of the columns have any non-zero values. 
Thus, changes to $\lambda$ are expected to produce only small changes to the estimates. 
Analysis of other penalties $\lambda \in \{0.1, 0.5, 5, 50\}$ (not shown) 
was consistent with this expectation.

Since the true model for peer and ego behavior is expected to be highly heterogeneous across very different URLs, we fit a separate model for each domain. This also facilitates a form of internal replication.
So for model $m$, the estimated propensity score for user $i$ being exposed to a URL $u$ from domain $d$ is
$$
\hat{e}_{d}(X_{iu}) =  \text{logit}^{-1} \big( X_{iu} \hat{\beta}_{d}  \big).
$$
This procedure is conducted for each of the models described below. (We suppress indication of which model $m$ is used from notation except where needed.)

The resulting estimated propensity scores can then be used in three closely related ways --- to construct weights for each unit, to match exposed and unexposed units, or to divide the sample into strata (i.e., subclasses). We use (post-)stratification (i.e., subclassfication) on the estimated propensity scores. Such stratification can also be regarded as form of nonparametric weighting or a form of matching, sometimes called ``blocking'' \citep{imbens_nonparametric_2004} or ``interval matching'' \citep{morgan_matching_2006}, that does not impose a particular ratio of treated to control units, as one-to-one matching methods do.
For very large data sets, such as the current study, stratification has computational advantages over matching, easily supports a much larger control group than treatment group, and the larger sample sizes afford using more strata than is otherwise common.\footnote{
Additionally, as discussed in Supplementary Information Section 2.3, the best available method for producing confidence intervals --- a multiway cluster-robust bootstrap strategy --- is inconsistent for nearest neighbor matching, rather than stratification, on propensity scores. For smaller data sets without this dependence structure, other recent developments, such as direct matching on many covariates could be preferable \citep{diamond_genetic_2013}. Post-stratification on propensity scores is not covered by some recent results by \citet{king_propensity_2016}, who argue against the use of propensity scores for matching.}
Together, these considerations motivated our use of propensity score stratification.

The boundaries of the strata are given by quantiles of the estimated propensity scores for each user--URL pair within each domain. For each domain, we use a number of strata $J$ proportional to the square-root of the number of exposed user--URL pairs, though any large number of strata yields similar results (see Supplementary Information Section 2.1).
So for each model $m$, domain $d$, and $j \in \{1, 2, ..., J\}$ we have an interval $\hat{Q}_{dj} \subset [0,1]$ of the scores between the $j-1$ to $j$th quantiles.\footnote{For some of the simpler models, discreteness in the estimate mean there are not J unique quantiles.} The strata-specific probability of sharing is estimated with a simple average of the outcomes for all the unexposed pairs in that strata
$$
\hat{p}^{(0)}_{dj} = \frac{1}{n^{(0)}_{dj}} \sum_{\langle i, u \rangle \in \text{C(d)}} Y_{iu} \mathbf{1}[\hat{e}_{d}(X_{iu}) \in \hat{Q}_{dj}]
$$
where $C(d)$ is the set of user--URL pairs in the NECG from domain $d$.
The estimate for a particular domain $d$ for model $m$ is an average of the estimates for each stratum weighted by the number of exposed pairs within that strata
$$
\hat{p}^{(0)}_{d} = \sum_{j = 1}^{J} \frac{n^{(1)}_{dj}}{n^{(1)}_{d}} \; \hat{p}^{(0)}_{dj}.
$$
Propensity score stratification thus results in weighting outcomes for unexposed individuals in the NECG according to the number of exposed units with similar propensity scores.
This process is illustrated with $J = 100$ strata for a single domain in Fig.~\ref{pss_process}.

Estimates from multiple domains are combined in the same way by weighting the estimate for each domain by the number of exposed pairs for that domain
$$
\hat{p}^{(0)} = \sum_{d} \frac{n^{(1)}_{d}}{n^{(1)}} \; \hat{p}^{(0)}_{d}.
$$
This weighted average of domain-specific estimates is then used to estimate the other quantities of interest (e.g., $\delta$, $RR$) in combination with the estimate of $p^{(0)}$ common to both the experimental and observational analyses.
\subsubsection{Efficiency}
Even under overlap and conditional unconfoundedness assumptions, this estimator will only achieve the asymptotic semiparametric efficiency bound for the ATT under stronger assumptions than some alternative methods \citep{hahn_role_1998}. If $e(X_{iu})$ were estimated nonparametrically, then weighting by the inverse of the estimated propensity scores would be efficient \cite{hirano_efficient_2003}, and similar arguments apply to post-stratification on the estimated propensity scores with a growing number of strata \citep{imbens_nonparametric_2004}. The sparse, high-dimensional covariates available here motivate instead using a parametric $L_2$-regularized logistic regression, at the cost of this asymptotic property. Alternative methods could combine this regularized propensity score model with an outcome model to achieve efficient estimation \citep{robins_semiparametric_1995,hahn_role_1998,imbens_nonparametric_2004,belloni_inference_2014,athey_efficient_2016,chernozhukov_double_2016}. In the present and similar settings, there are reasons to think the potential variance reductions from modeling the outcome are small. In particular, with a rare ($<$0.2\%) binary outcome, there is little information in the outcome for such methods to exploit. We selected the present methods in part because of their appealing computational properties forlarge data sets.
\subsection{Sets of covariates}
There are numerous variables available for the propensity score model.\footnote{For example, an analyst could construct the \emph{individual--term matrix} counting all the words used by each individual in their Facebook communications; each of these thousands of variables could be used as a covariate.}
It is not possible or desirable to include all the variables that an analyst could construct because of the work involved in defining variables, the costs of increasing dimensionality for precision, and computational challenges in using all of them in an analysis. Furthermore, many situations may require that investigators decide in advance what variables are worth measuring. In both of these cases, it is standard practice to use theory and other domain knowledge to select variables. In the case of peer effects in URL sharing, the analyst would select variables believed to be related to causes of sharing a URL and to be associated with network structure (i.e., peer and ego variables are associated because of homophily, common external causes, and prior influence). This is not to say that the analyst must think each variable is a likely cause of sharing behaviors, but simply that they are causes of sharing behaviors \emph{or} are descendants of these causes.

\renewcommand{\arraystretch}{1.3}

\begin{table}[htp]
\caption{Variables included in models predicting exposure. The final column indicates which base model specification include that variable. Some variables are transformed and/or contribute multiple inputs (columns) to a model. \dag: Includes untransformed and squared terms, $x$ and $x^2$; *: Transformed with $\log(x + 1)$; \ddag: Includes binary indicator, $1{\{x > 0\}}$. All other variables are untransformed; if categorical, one indicator (dummy) for each level is included in the model matrix.}
\begin{center}
\small{
\begin{tabular}{p{2.3cm} p{3.6cm} p{6cm} r r}
\hline
Category & Name & Description & Columns & Models\\ \hline
Demographics 
& Age$^\dag$ & As indicated on profile & 2 & $\mn{A}, \mn{D}$\\
& Gender & Indicated or inferred: female, male, or unknown & 2 & $\mn{A}, \mn{D}$ \\ 
\hline
Facebook 
& Friend count & Number of extant friendships& 1 & $\mn{A}$ \\
& Friend initiation & Number and proportion of extant friendships initiated & 2 &$\mn{A}$ \\
& Tenure$^*$ & Days since registration of account & 1 &$\mn{A}$ \\
& Profile picture & Whether the user has a profile picture  & 1 & $\mn{A}$ \\
& Visitation freq. & Days active in prior 30, 91, and 182 day periods  & 3 & $\mn{A}$ \\ 
\hline
Communication 
& Action count$^*$ & Number of posts (including URLs), comments, and likes in a prior one month period & 1 & $\mn{A}$ \\
& Post count$^*$ & Number of posts (including URLs) in a prior one month period & 1 & $\mn{A}$ \\
& Comment count$^*$ & Number of comments on posts in a prior one month period & 1 & $\mn{A}$ \\
& Like count$^*$ & Number of posts and comments ``liked'' in a prior one month period & 1 & $\mn{A}$ \\
\hline
Link sharing 
& Shares$^{* \ddag}$ & Number of URLs shared in a one month period  & 2 & $\mn{A}$ \\
& Unique domains$^*$ & Number of unique domains of URLs shared in a six month period  & 1 &$\mn{A}$ \\ 
& Same domain shares$^{* \ddag}$  & Number of times shared any URL with the same domain as outcome URL in six month period& 2 & $\mn{s}$\\
& Other domain shares$^*$  & Number of times shared any URL in six month period for each of the other domains & 3,703 & $\mn{M}$\\	
                     \hline
\end{tabular}
}
\end{center}
\label{available_variables}
\end{table}%

Table \ref{available_variables} lists the variables we computed for use as covariates. These variables are each included in at least one of model specifications, which are designed to correspond to selections of variables that an analyst might make and to evaluate the contribution of different sets of variables to bias reduction. Model $\mn{A}$ includes all of the base variables. This model is expected to have the largest potential for bias reduction but to also suffer from increased sampling variance. In other settings, many of these variables might not be available to analysts. Model $\mn{D}$ includes demographic variables only. At least some of these variables, or similar measurements, would likely be available in many other settings. These are all expected to be associated with consuming content from particular sources. $\mn{D}$ can also be seen as a relatively minimal convenience selection of covariates.

We consider two additional sets of predictors that can be combined with these sets of covariates. First, we expected that, by virtue of serving as measures of a user's latent interest in and likelihood of independently encountering a URL, variables describing prior interactions with related URLs could result in substantial bias reduction. In particular, for some user--URL pair $iu$, let \emph{same domain shares} count the number of URLs that $i$ shared in the six months prior to the experiment that have the same domain name as $u$. Models that add this variable are indicated with $\mn{s}$; for example, Model $\mn{As}$ adds same domain shares to Model $\mn{A}$. This allows for straightforward evaluation of the consequences of using this variable to the observational analysis. 

We regard same domain shares as an example of more specific information about related prior behaviors. In some cases, such information will be available to analysts. In other cases, this information may not be available, or the related behaviors may not be sufficiently common to be useful. In particular, if the focal behavior is new (e.g., a new product launch) or only recently popular, then this information may be limited. In the present case, very few users may have shared any URLs from a particular domain during the prior six months; that is, same domain shares can be 0 for most or all users for some domains.

For this reason, we also evaluate models that include the number of times a user shared URLs from each of the other 3,703 domain names; we indicate the presence of these predictors with $\mn{M}$, as this corresponds to the addition of a large sparse matrix of (log-transformed) counts. These models have important similarities with the use of low-rank matrix decomposition methods in, e.g., recommendation systems: the $L_2$ penalty results in shrinking larger principal components of the training data less, where many matrix decomposition methods would simply select a small number of components to use to represent the tastes of individuals \citep[][\S 3.4.1]{hastie_elements_2008}.

\subsection{Comparisons of estimators}
To evaluate observational estimators of the relative risk $RR$ and risk difference (or ATT) $\delta$, we use the NECG, as described above, to produce estimates of $p^{(0)}$ that make no use of the control group from the randomized experiment.
Recall that the experimental and observational estimates of ${p}^{(1)}$ are identical, as they are both the proportion of exposed user--URL pairs that resulted in sharing; thus, all discrepancies are due to differences in estimating $p^{(0)}$.

We compute the discrepancy between each of the resulting observational estimates and the experimental estimates. Our focus is primarily on estimates of the relative risk $RR$. We also consider the risk difference $\delta$ (i.e., the average treatment effect on the treated, ATT). 
For each observational estimator $m$, we have two estimators, $\widehat{RR}_m$ and $\hat{\delta}_m$. We generally take the experimental estimates as the gold standard --- as unbiased for the causal estimand of interest. This motivates the description of these discrepancies as estimates of bias.

For the relative risk, we can compute the absolute discrepancy in the estimates, $
\widehat{RR}_m - \widehat{RR}_{\text{exp}}$.
To put this is relative terms, we can compute the relative percent bias in the relative risk:
$$
100 \frac{\widehat{RR}_m - \widehat{RR}_{\text{exp}}}{\widehat{RR}_{\text{exp}}}.
$$

For the risk difference, we can also compute absolute and percent bias, similarly to above.
Since the risk difference $\delta = p^{(1)} - p^{(0)}$ is bounded from above by $p^{(1)}$ (i.e. when the behavior cannot occur without exposure), the maximum possible overestimate of $\delta$ is too large by $p^{(0)}$. Thus, we can also characterize error in terms of this maximum possible overestimate, percent bias of the maximum possible overestimate:
$$
100 \frac{\hat{\delta}_m - \hat{\delta}_{\text{exp}}}{\hat{p}^{(0)}}
$$
where we assume $\hat{\delta}_m \geq \hat{\delta}_{\text{exp}}$.

To account for dependence among observations of the same user or same URL,
all confidence intervals reported in this paper are 95\% standard bootstrap confidence intervals robust to dependence among repeated observations of both users and URLs \citep{owen_bootstrapping_2012} and accounting for the sampling error in both the experimental and observational estimates; see Supplementary Information Sections 2.3.
\section{Results}
\label{sec:results}
On average a user exposed to a peer sharing a URL (i.e., a user--URL pair in the \emph{feed} condition) goes on to share that URL 0.130\% of the time,
while a user who was not exposed to a URL because that user--URL pair was randomly assigned to the  \emph{no feed} condition goes on to share that URL 0.019\% of the time.
That is, exposure to a peer sharing a URL causes sharing for $\hat{\delta}_{\text{exp}} = $ 
0.111\% of pairs (CI  = [0.109, 0.114]), and users are $\widehat{RR}_{\text{exp}} = $ 6.8 times as likely to share a URL in the \emph{feed} condition 
compared to those in the \emph{no feed} condition (CI = [6.5, 7.0]). 
These are the experimental estimates of peer effects to which we compare observational estimates.

\begin{sidewaysfigure}[hp]
\begin{center}
\subfloat[]{\includegraphics[width=\overallsubfigwidth]{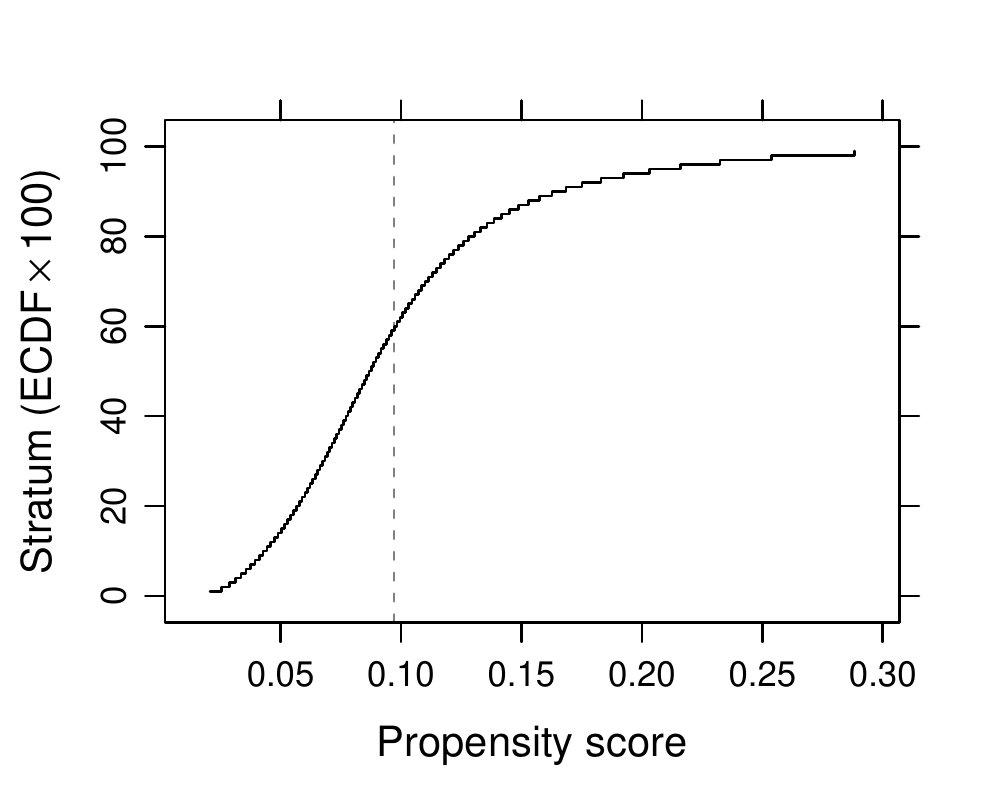}} 
\subfloat[]{\includegraphics[width=\overallsubfigwidth]{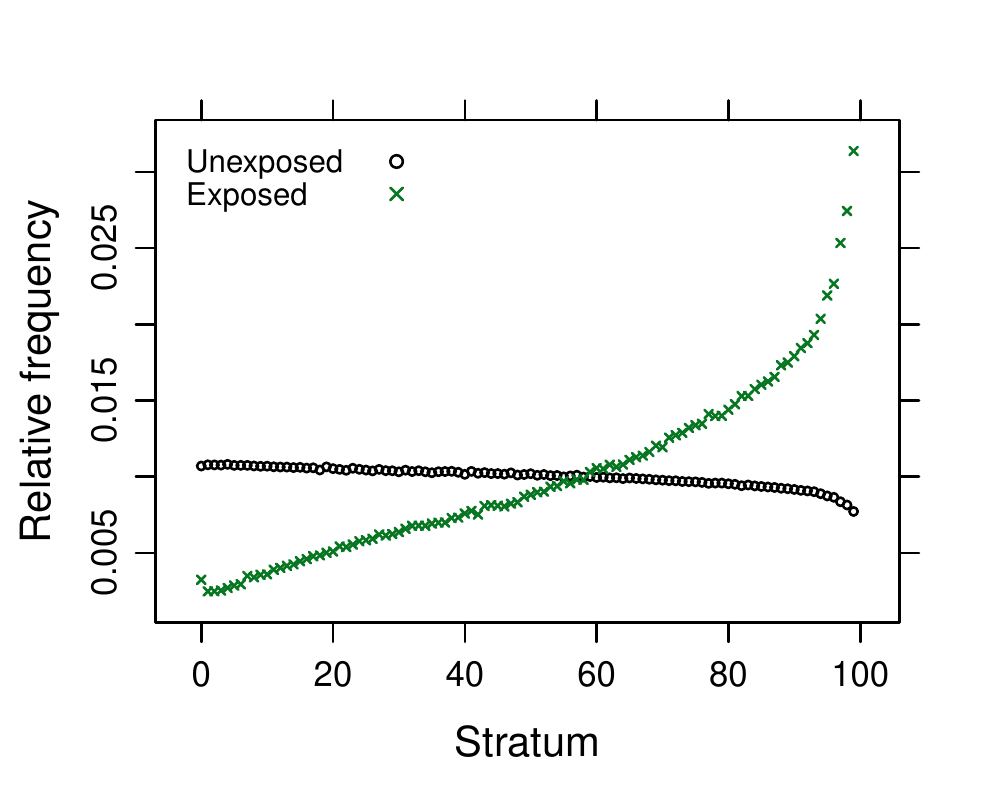}} 
\subfloat[]{\includegraphics[width=\overallsubfigwidth]{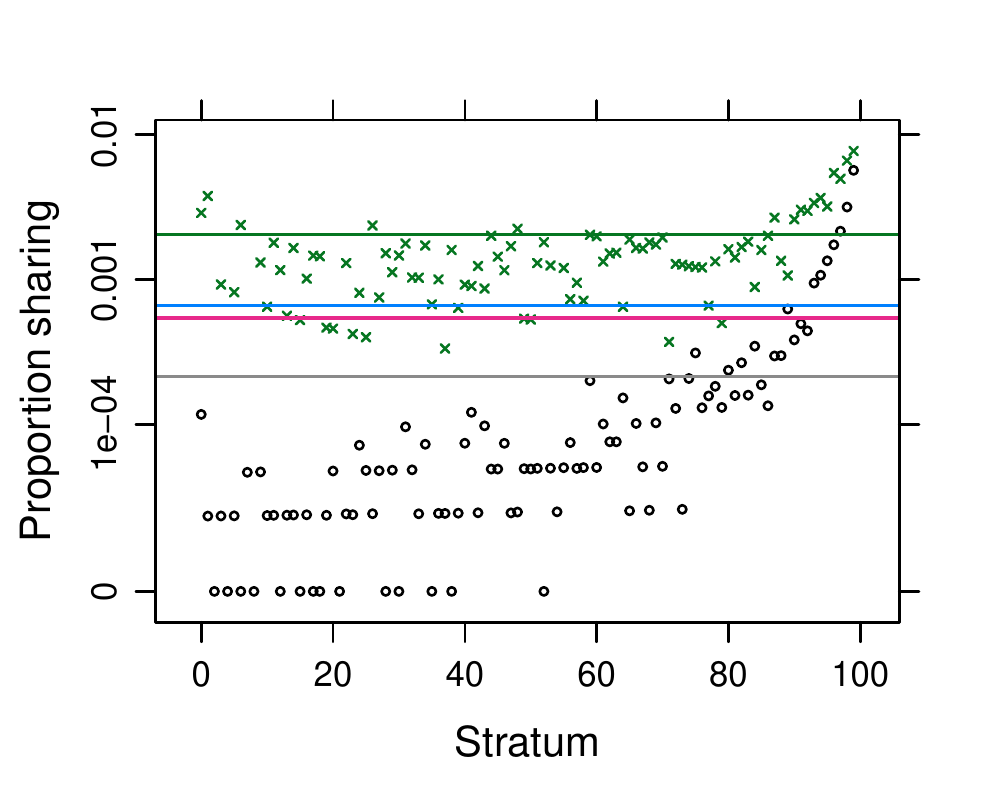}}
\end{center}
\caption{
{\bf High-dimensional propensity score modeling and stratification for the sharing probabilities.} Illustration for a popular domain (\text{www.nytimes.com}) with 100 strata.
A propensity score model is fit predicting the probability a given user--URL pair is exposed (i.e. the user is exposed to that URL by a peer sharing it), using thousands of covariates (corresponding to the $\mn{AMs}$ model).
Observations are mapped to 100 strata (i.e., percentiles) based on the ECDF of the estimated propensity scores of exposed observations and unexposed observations in the NECG (A); there is variation around the grand mean (dashed line), and this variation is greater than for the models with fewer covariates (see Supplementary Materials Section 2.4).
As expected, exposed pairs are more common among higher strata (B).
The probability of sharing a URL is greater for higher strata, both for exposed and unexposed users (C).
The naive observational estimate of ${p}^{(0)}$ (grey line) weights the probability of sharing for the unexposed in each stratum by their relative frequencies (as does ignoring the stratification). Propensity score stratification instead weights sharing probabilities by the relative frequency of exposed pairs.
This estimate of ${p}^{(0)}$ (magenta line) is much closer to the gold-standard, experimental estimate of ${p}^{(0)}$ (blue line). 
The estimate of ${p}^{(1)}$ common to both the experimental and observational analysis is shown is superimposed (green line).
}
\label{pss_process}
\end{sidewaysfigure}

\begin{sidewaysfigure}[hp]
\begin{center}
\includegraphics[width=1\textwidth]{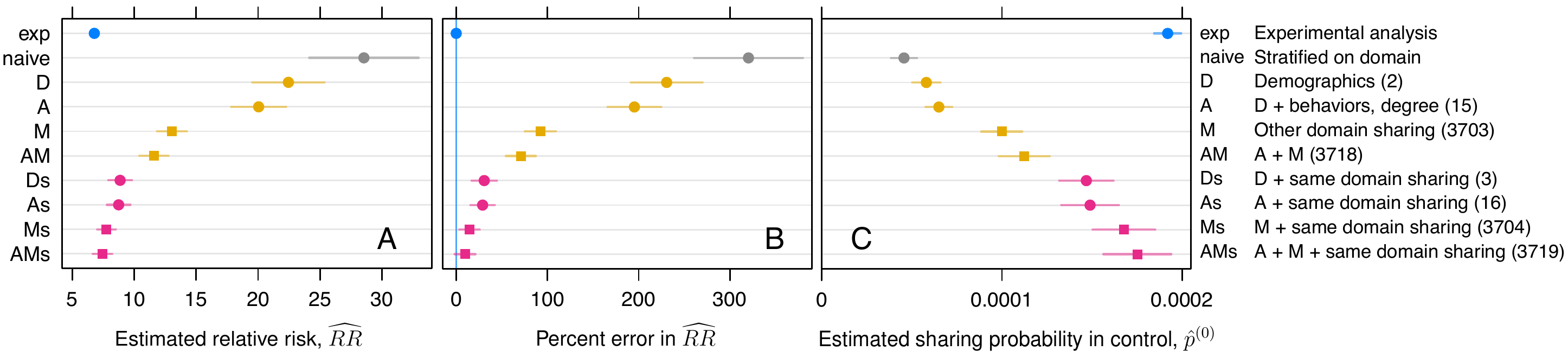}
\end{center}
\caption{
{\bf Comparison of experimental and observational estimates of peer effects.} 
(A) The experiment estimates that users are 6.8 times as likely to share when exposed to a peer sharing, while the observational point estimates are larger. 
(B) Treating the experimental estimate as the truth, the naive observational estimate overestimates peer effects by 320\%. This bias is substantially reduced by adjusting for prior same domain sharing (magenta) and prior sharing for 3,703 other domains (squares).
(C) All discrepancies in the estimates of relative risk are due to underestimating $p^{(0)}$ when using observational data.
Error bars are 95\% confidence intervals. Brief descriptions of the estimators with number of covariates in parentheses are shown for reference.
}
\label{pss_overall}
\end{sidewaysfigure}

\renewcommand{\arraystretch}{3}
\begin{table}[htp]
\caption{
{\bf Comparison of experimental and observational estimates of peer effects.} 
Estimates of the probability of sharing if not exposed ${p}^{(0)}$, relative risk ($RR$), and the risk difference ($\delta$) for each model with 95\% bootstrap standard confidence intervals in brackets.}
\begin{center}
\begin{tabular}{ r  c  c  c  c }
\hline
Model & $\hat{p}^{(0)}$ & $\widehat{RR}$ & $\hat{\delta}$ \\ 
  \hline
AMs & \shortstack{ 1.751e-04 \\ \relax [1.563e-04, 1.940e-04] } & \shortstack{ 7.44 \\ \relax [6.65, 8.23] } & \shortstack{ 1.128e-03 \\ \relax [1.097e-03, 1.159e-03] } \\ 
  Ms & \shortstack{ 1.677e-04 \\ \relax [1.503e-04, 1.851e-04] } & \shortstack{ 7.77 \\ \relax [7.02, 8.52] } & \shortstack{ 1.135e-03 \\ \relax [1.109e-03, 1.162e-03] } \\ 
  AM & \shortstack{ 1.124e-04 \\ \relax [9.828e-05, 1.265e-04] } & \shortstack{ 11.59 \\ \relax [10.42, 12.77] } & \shortstack{ 1.191e-03 \\ \relax [1.163e-03, 1.218e-03] } \\ 
  M & \shortstack{ 9.989e-05 \\ \relax [8.855e-05, 1.112e-04] } & \shortstack{ 13.05 \\ \relax [11.84, 14.25] } & \shortstack{ 1.203e-03 \\ \relax [1.179e-03, 1.228e-03] } \\ 
  As & \shortstack{ 1.489e-04 \\ \relax [1.329e-04, 1.649e-04] } & \shortstack{ 8.75 \\ \relax [7.80, 9.70] } & \shortstack{ 1.154e-03 \\ \relax [1.127e-03, 1.181e-03] } \\ 
  Ds & \shortstack{ 1.469e-04 \\ \relax [1.318e-04, 1.619e-04] } & \shortstack{ 8.87 \\ \relax [7.92, 9.82] } & \shortstack{ 1.156e-03 \\ \relax [1.128e-03, 1.183e-03] } \\ 
  A & \shortstack{ 6.501e-05 \\ \relax [5.761e-05, 7.241e-05] } & \shortstack{ 20.04 \\ \relax [17.82, 22.27] } & \shortstack{ 1.238e-03 \\ \relax [1.214e-03, 1.263e-03] } \\ 
  D & \shortstack{ 5.806e-05 \\ \relax [5.018e-05, 6.593e-05] } & \shortstack{ 22.45 \\ \relax [19.53, 25.36] } & \shortstack{ 1.245e-03 \\ \relax [1.221e-03, 1.269e-03] } \\ 
  naive & \shortstack{ 4.567e-05 \\ \relax [3.842e-05, 5.291e-05] } & \shortstack{ 28.54 \\ \relax [24.12, 32.95] } & \shortstack{ 1.257e-03 \\ \relax [1.233e-03, 1.282e-03] } \\ 
  exp & \shortstack{ 1.920e-04 \\ \relax [1.845e-04, 1.995e-04] } & \shortstack{ 6.79 \\ \relax [6.54, 7.04] } & \shortstack{ 1.111e-03 \\ \relax [1.086e-03, 1.136e-03] } \\ 
   \hline

\end{tabular}
\end{center}
\label{table:overall_ests}
\end{table}

The naive observational analysis, which makes no adjustment for observed covariates, 
concludes that exposure makes uses sharing 28.5 times as likely (CI = [24.1, 33.0]). 
That is, it overestimates peer effects by 320\% (Fig.~\ref{pss_overall}A-B). These results are also displayed in Table \ref{table:overall_ests}.
Because $p^{(0)}$ is bounded from below by zero (Fig.~\ref{pss_overall}C), 
$\hat{p}^{(1)} =$ 0.00130 is the maximum possible estimate of $\delta$ (i.e. if all sharing is attributed to peer effects). The naive observational analysis yields $\hat{\delta}_\text{naive} =$ 0.00126 thus overstates $\delta$ by 76\% of this maximum; in this sense it barely improves on attributing all sharing to peer effects.

We then evaluate observational estimators that use propensity score models \citep{rosenbaum_central_1983}, which have been used in many of the most credible observational studies of peer effects \citep{aral_distinguishing_2009}, estimated with $L_2$-penalized logistic regression and varied sets of covariates; see Fig. \ref{pss_process} and Methods. Only adjusting for demographic ($\mn{D}$) or basic individual-level covariates ($\mn{A}$) resulted in similarly large overestimates by 231\% and 195\% (Fig.~\ref{pss_overall}B).

One covariate measuring prior, highly-related behaviors was of particular interest and expected to substantially reduce bias \emph{a priori}: For each user--domain-name pair, the \emph{prior same domain sharing} variable counts how many URLs from that domain name that user shared in the 6 month pre-experiment period (e.g., for the URL \url{http://www.cnn.com/article_x}, this is how many URLs the user shared with the domain \url{www.cnn.com}). Estimators that additionally adjusted for prior same domain sharing ($\mn{Ds}$ and $\mn{As}$) conclude that exposure makes sharing 8.9 (CI = [7.9, 9.8]) or 8.8 (CI = [7.8, 9.7]) times as likely --- both overestimating the relative risk by only 31\% and 29\%.

Rather than selecting a single measure of prior sharing (e.g., from the same domain) \emph{a priori}, we can use a high-dimensional $L_2$-penalized model including measures of prior sharing for all other 3,703 included domains. To the extent that these measures are correlated (e.g., they reflect interest in a common topic), the $L_2$ penalty in the propensity score model can be interpreted as penalizing components common to many and popular domains less than idiosyncratic components, as in widely-used methods for product and item recommendation.
The model with all 3,719 covariates ($\mn{AMs}$),
including prior sharing for all domains, eliminated 97\% of the naive estimator's bias for the relative risk.
This corresponds to concluding that exposure makes sharing
7.4 times as likely (CI~=~[6.7, 8.2]).
This point estimate is less than 10\% larger than and statistically equivalent with the experimental estimate of 6.8 ($p >$ 0.05) .
Instead of supplementing an \emph{a priori} related covariate, as in this full model, such high-dimensional adjustment might be used as a substitute when no such covariate is measured. In the absence of prior same domain sharing, the sparse matrix of prior sharing for all other domains reduces bias (i.e., $\mn{AM}$ compared with $\mn{A}$, or $\mn{AM}$ compared with the naive estimator), but does not fully substitute for also adjusting for prior same domain sharing (i.e., $\mn{AM}$ compared with $\mn{As}$); see Supplementary Information Table S1.

\subsection{Heterogeneity by prior popularity}
\begin{figure*}[hp]
\begin{center}
\includegraphics[width=.75\columnwidth]{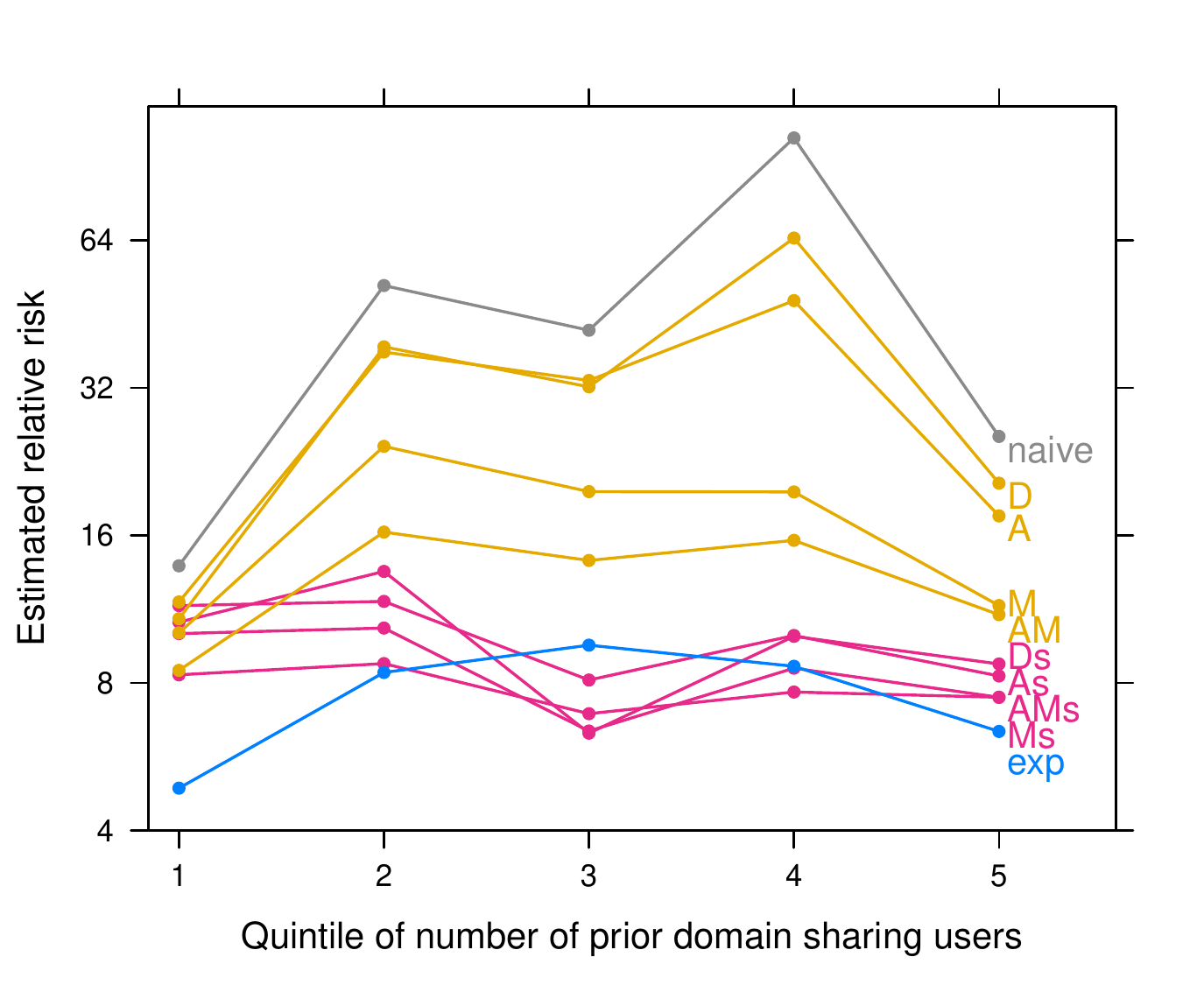}
 \caption{{\bf Estimated relative risk of sharing as a function of a domain's prior popularity.} Popularity is given in terms of quintiles of the number of unique users sharing URLs from the domain. For previously unpopular domains, all of the observational analyses are similarly biased. As with the overall results in Fig.~\ref{pss_overall}, estimates for URLs from domains that were popular prior to the experiment exhibit more bias reduction when adjusting for prior same domain sharing and/or prior sharing from other domains.}
\label{pss_domain_pop_error}
\end{center}
\end{figure*}

These results show that striking levels of bias reduction are possible when pooling across all domains in our dataset, which aggregates over 11.5 million distinct behaviors (i.e., sharing a particular URL). Much of this bias reduction results from adjusting for prior same domain sharing, and over 58\% of the exposures come from the top 10\% of the domains, such that the bias reduction obtained may largely be due to well-established, highly popular classes of behaviors. However, much of the peer effects literature focuses on the early stages of the spread of new behaviors [e.g., adoption of new products, opinions, or adaptive behaviors \citep{iyengar_opinion_2011,ugander_structural_2012,myers_information_2012,aplin_experimentally_2015}].
How much bias reduction is possible for less common or newly popular classes of behaviors?  We analyze the estimates of peer effects for each of the domains with respect to how popular those domains are for the 6 months prior to the study (3,704 domains have at least one prior-sharing user in the sample). Simple subgroup analyses on groups of domains defined by quintiles of prior popularity shows that the observational estimators remain significantly biased for the least previously popular domains (Fig.~\ref{pss_domain_pop_error}): 
For these previously unpopular domains (i.e., bottom quintile), all of the observational estimates are substantively similar, but as prior popularity increases, substantial differences among the estimates emerge. The domains that were most popular before the study are also popular during the study (Supplementary Information Section 3.3).  Thus, much of bias reduction for the overall estimates (Fig.~\ref{pss_overall}) can be attributed to bias reduction for the domains with the greatest prior popularity.

\section{Discussion}
The study of peer effects, while central to the social sciences, has been limited by biases that have been difficult to quantify.
Field experiments are a promising solution to these problems, but restricting scientific inquiry about peer effects to questions answerable with field experiments would severely limit research in this area \citep{christakis_social_2013}:
few organizations are able to run experiments with sufficient statistical power to precisely estimate peer effects, it is often impractical to run real-world experiments, 
and it is not possible to run experiments to retrospectively study the contribution of peer effects to important events. We conducted the first evaluation of observational studies of peer effects to use a large field experiment as its comparison point. Treating experimental results as the ``gold standard'', we find substantial variation in how well observational estimators perform: analyses that only adjust for a small set of common demographic and variables suffer from nearly as much bias as unadjusted, naive estimates; but estimates adjusting for a relevant prior behavior selected \emph{a priori} or thousands of potentially relevant behaviors are able to remove the majority of bias. We note that this bias reduction depends on the presence of meaningful variation in this prior behavior; newer classes of behavior (i.e., sharing URLs from domains not previously popular) do not exhibit this substantial bias reduction. With or without this measure selected \emph{a priori}, high-dimensional adjustment using thousands of potentially relevant prior behaviors reduced bias.

These results show how causal inference with routinely collected data can be improved through both substantive knowledge and high-dimensional statistical learning techniques.
Scientists studying peer effects can use existing knowledge to inform the selection and construction of relevant measures for adjustment.
Our specific methods, granular stratification on propensity scores estimated using high-dimensional penalized regression, are directly applicable to a number of data sets where measures of prior behaviors like those used in this study are available, including online communication behaviors, purchase history data (i.e., ``scanner data''), history of drug prescriptions by individual doctors \citep{iyengar_opinion_2011}, or data from passive transponder tags on animals \citep{allen_network_2013,aplin_experimentally_2015}.
The success of high-dimensional adjustment here should encourage scientists to measure a larger number of prior behaviors and to employ modern statistical learning techniques with large data sets for causal inference, rather than just description and prediction \citep{varian_causal_2016}.

This study should not be understood as validating all observational studies of peer effects that adjust for prior behaviors.
First, while this experiment includes millions of specific behaviors, they are similar to each other in important ways, and spread via the same communication platform. 
Other behaviors may differ in their prevalence, size of peer effects, costs of adoption, and the time-scales at which they occur. For example, the posited processes producing correlations of obesity in social networks \citep{christakis_spread_2007} occur over a long period of time and are presumably the result of peer effects in many contributing behaviors (e.g., diet, exercise).
Second, evaluating observational methods requires considering how the resulting estimates are used by scientists and decision-makers. Many of the best-performing observational estimators in this study still overestimated peer effects. How problematic such bias is depends on the specific decisions made with such estimates  (e.g., retaining or rejecting a theory, making a change to a marketing strategy). Qualitatively, an exposed individual being 7.4 or 8.8 times as likely to perform a behavior is similar to 6.8 times as likely, but such a difference could matter for computing and comparing return on investment from, e.g., public health marketing campaigns \citep[cf.][]{lewis_unfavorable_2015}.

\small
\singlespacing
\bibliographystyle{apalike}
\bibliography{peerinfluence,causalinference,other}

\begin{thebibliography}{}

\bibitem[Allcott and Gentzkow, 2017]{allcott_social_2017}
Allcott, H. and Gentzkow, M. (2017).
\newblock Social media and fake news in the 2016 election.
\newblock {\em Journal of Economic Perspectives}, 31(2):211--236.

\bibitem[Allen et~al., 2013]{allen_network_2013}
Allen, J., Weinrich, M., Hoppitt, W., and Rendell, L. (2013).
\newblock Network-based diffusion analysis reveals cultural transmission of
  lobtail feeding in humpback whales.
\newblock {\em Science}, 340(6131):485--488.

\bibitem[Angrist, 2014]{angrist_perils_2014}
Angrist, J.~D. (2014).
\newblock The perils of peer effects.
\newblock {\em Labour Economics}, 30:98--108.

\bibitem[Aplin et~al., 2015]{aplin_experimentally_2015}
Aplin, L.~M., Farine, D.~R., Morand-Ferron, J., Cockburn, A., Thornton, A., and
  Sheldon, B.~C. (2015).
\newblock Experimentally induced innovations lead to persistent culture via
  conformity in wild birds.
\newblock {\em Nature}, 518(7540):538--541.

\bibitem[Aral et~al., 2009]{aral_distinguishing_2009}
Aral, S., Muchnik, L., and Sundararajan, A. (2009).
\newblock Distinguishing influence-based contagion from homophily-driven
  diffusion in dynamic networks.
\newblock {\em Proceedings of the National Academy of Sciences},
  106(51):21544--21549.

\bibitem[Aral and Walker, 2011]{aral_creating_2011}
Aral, S. and Walker, D. (2011).
\newblock Creating social contagion through viral product design: A randomized
  trial of peer influence in networks.
\newblock {\em Management Science}, 57(9):1623--1639.

\bibitem[Asch, 1956]{asch_studies_1956}
Asch, S.~E. (1956).
\newblock Studies of independence and conformity: I. {A} minority of one
  against a unanimous majority.
\newblock {\em Psychological Monographs: General and Applied}, 70(9):1--70.

\bibitem[Athey et~al., 2016]{athey_efficient_2016}
Athey, S., Imbens, G.~W., Wager, S., et~al. (2016).
\newblock Efficient inference of average treatment effects in high dimensions
  via approximate residual balancing.
\newblock https://arxiv.org/abs/1604.07125.

\bibitem[Bakshy et~al., 2012a]{bakshy_social_2012}
Bakshy, E., Eckles, D., Yan, R., and Rosenn, I. (2012a).
\newblock Social influence in social advertising: Evidence from field
  experiments.
\newblock In {\em Proceedings of the {ACM} conference on Electronic Commerce}.
  {ACM}.

\bibitem[Bakshy et~al., 2011]{bakshy_everyones_2011}
Bakshy, E., Hofman, J.~M., Mason, W.~A., and Watts, D.~J. (2011).
\newblock Everyone's an influencer: {Q}uantifying influence on {T}witter.
\newblock In {\em Proceedings of the fourth {ACM} international conference on
  Web search and data mining}, {WSDM} '11, pages 65--74. {ACM}.

\bibitem[Bakshy et~al., 2015]{bakshy_exposure_2015}
Bakshy, E., Messing, S., and Adamic, L.~A. (2015).
\newblock Exposure to ideologically diverse news and opinion on {F}acebook.
\newblock {\em Science}, 348(6239):1130--1132.

\bibitem[Bakshy et~al., 2012b]{bakshy_role_2011}
Bakshy, E., Rosenn, I., Marlow, C., and Adamic, L. (2012b).
\newblock The role of social networks in information diffusion.
\newblock In {\em Proceedings of the 21st international conference on World
  Wide Web}, WWW '12, pages 519--528. ACM.

\bibitem[Banerjee et~al., 2013]{banerjee_diffusion_2013}
Banerjee, A., Chandrasekhar, A.~G., Duflo, E., and Jackson, M.~O. (2013).
\newblock The diffusion of microfinance.
\newblock {\em Science}, 341(6144):1236498.

\bibitem[Barber{\'a} et~al., 2015]{barbera_tweeting_2015}
Barber{\'a}, P., Jost, J.~T., Nagler, J., Tucker, J.~A., and Bonneau, R.
  (2015).
\newblock Tweeting from left to right: Is online political communication more
  than an echo chamber?
\newblock {\em Psychological Science}, 26(10):1531--1542.

\bibitem[Belloni et~al., 2014]{belloni_inference_2014}
Belloni, A., Chernozhukov, V., and Hansen, C. (2014).
\newblock Inference on treatment effects after selection among high-dimensional
  controls.
\newblock {\em The Review of Economic Studies}, 81(2):608--650.

\bibitem[Berger, 2011]{berger_arousal_2011}
Berger, J. (2011).
\newblock Arousal increases social transmission of information.
\newblock {\em Psychological Science}, 22(7):891--893.

\bibitem[Blume, 1995]{blume_statistical_1995}
Blume, L.~E. (1995).
\newblock The statistical mechanics of best-response strategy revision.
\newblock {\em Games and Economic Behavior}, 11(2):111--145.

\bibitem[Bond et~al., 2012]{bond_massive_2012}
Bond, R.~M., Fariss, C.~J., Jones, J.~J., Kramer, A. D.~I., Marlow, C., Settle,
  J.~E., and Fowler, J.~H. (2012).
\newblock A 61-million-person experiment in social influence and political
  mobilization.
\newblock {\em Nature}, 489(7415):295--298.

\bibitem[Cai et~al., 2015]{cai_social_2015}
Cai, J., De~Janvry, A., and Sadoulet, E. (2015).
\newblock Social networks and the decision to insure.
\newblock {\em American Economic Journal: Applied Economics}, 7(2):81--108.

\bibitem[Card and Giuliano, 2013]{card_peer_2013}
Card, D. and Giuliano, L. (2013).
\newblock Peer effects and multiple equilibria in the risky behavior of
  friends.
\newblock {\em The Review of Economics and Statistics}, 95(4):1130--1149.

\bibitem[Carrell et~al., 2009]{carrell_does_2009}
Carrell, S.~E., Fullerton, R.~L., and West, J.~E. (2009).
\newblock Does your cohort matter? {M}easuring peer effects in college
  achievement.
\newblock {\em Journal of Labor Economics}, 27(3):439--464.

\bibitem[Centola, 2010]{centola_spread_2010}
Centola, D. (2010).
\newblock The spread of behavior in an online social network experiment.
\newblock {\em Science}, 329(5996):1194--1197.

\bibitem[Centola and Macy, 2007]{centola_complex_2007}
Centola, D. and Macy, M. (2007).
\newblock Complex contagions and the weakness of long ties.
\newblock {\em American Journal of Sociology}, 113:702--734.

\bibitem[Cheng et~al., 2014]{cheng_can_2014}
Cheng, J., Adamic, L., Dow, P.~A., Kleinberg, J.~M., and Leskovec, J. (2014).
\newblock Can cascades be predicted?
\newblock In {\em Proceedings of the 23rd international conference on World
  wide web}, pages 925--936. ACM.

\bibitem[Chernozhukov et~al., 2016]{chernozhukov_double_2016}
Chernozhukov, V., Chetverikov, D., Demirer, M., Duflo, E., Hansen, C., et~al.
  (2016).
\newblock Double machine learning for treatment and causal parameters.
\newblock https://arxiv.org/abs/1608.00060.

\bibitem[Christakis and Fowler, 2007]{christakis_spread_2007}
Christakis, N.~A. and Fowler, J.~H. (2007).
\newblock The spread of obesity in a large social network over 32 years.
\newblock {\em N Engl J Med}, 357(4):370--379.

\bibitem[Christakis and Fowler, 2013]{christakis_social_2013}
Christakis, N.~A. and Fowler, J.~H. (2013).
\newblock Social contagion theory: {E}xamining dynamic social networks and
  human behavior.
\newblock {\em Statistics in Medicine}, 32(4):556--577.

\bibitem[{Cohen-Cole} and Fletcher, 2008]{cohen-cole_detecting_2008}
{Cohen-Cole}, E. and Fletcher, J.~M. (2008).
\newblock Detecting implausible social network effects in acne, height, and
  headaches: Longitudinal analysis.
\newblock {\em {BMJ}}, 337(2):2533--2533.

\bibitem[Cook et~al., 2008]{cook_three_2008}
Cook, T.~D., Shadish, W.~R., and Wong, V.~C. (2008).
\newblock Three conditions under which experiments and observational studies
  produce comparable causal estimates: New findings from within‐study
  comparisons.
\newblock {\em Journal of Policy Analysis and Management}, 27(4):724--750.

\bibitem[Dehejia, 2005]{dehejia_practical_2005}
Dehejia, R. (2005).
\newblock Practical propensity score matching: A reply to {S}mith and {T}odd.
\newblock {\em Journal of Econometrics}, 125(1):355--364.

\bibitem[Dehejia and Wahba, 1999]{dehejia_causal_1999}
Dehejia, R.~H. and Wahba, S. (1999).
\newblock Causal effects in nonexperimental studies: Reevaluating the
  evaluation of training programs.
\newblock {\em Journal of the American Statistical Association},
  94(448):1053--1062.

\bibitem[Dehejia and Wahba, 2002]{dehejia_propensity_2002}
Dehejia, R.~H. and Wahba, S. (2002).
\newblock Propensity score-matching methods for nonexperimental causal studies.
\newblock {\em Review of Economics and Statistics}, 84(1):151--161.

\bibitem[Diamond and Sekhon, 2013]{diamond_genetic_2013}
Diamond, A. and Sekhon, J.~S. (2013).
\newblock Genetic matching for estimating causal effects: A general
  multivariate matching method for achieving balance in observational studies.
\newblock {\em Review of Economics and Statistics}, 95(3):932--945.

\bibitem[Diaz and Handa, 2006]{diaz_assessment_2006}
Diaz, J.~J. and Handa, S. (2006).
\newblock An assessment of propensity score matching as a nonexperimental
  impact estimator: Evidence from mexico'€™s {PROGRESA} program.
\newblock {\em Journal of Human Resources}, 41(2):319--345.

\bibitem[Ding and Miratrix, 2015]{ding_adjust_2015}
Ding, P. and Miratrix, L.~W. (2015).
\newblock To adjust or not to adjust? {S}ensitivity analysis of {M}-bias and
  butterfly-bias.
\newblock {\em Journal of Causal Inference}, 3(1):41--57.

\bibitem[Eckles et~al., 2016]{eckles_estimating_2016}
Eckles, D., Kizilcec, R.~F., and Bakshy, E. (2016).
\newblock Estimating peer effects in networks with peer encouragement designs.
\newblock {\em Proceedings of the National Academy of Sciences},
  113(27):7316--7322.

\bibitem[Firth et~al., 2016]{firth_pathways_2016}
Firth, J.~A., Sheldon, B.~C., and Farine, D.~R. (2016).
\newblock Pathways of information transmission among wild songbirds follow
  experimentally imposed changes in social foraging structure.
\newblock {\em Biology Letters}, 12(6):20160144.

\bibitem[Flaxman et~al., 2016]{flaxman_filter_2016}
Flaxman, S., Goel, S., and Rao, J.~M. (2016).
\newblock Filter bubbles, echo chambers, and online news consumption.
\newblock {\em Public Opinion Quarterly}, 80(1):298--320.

\bibitem[Fortin and Yazbeck, 2015]{fortin_peer_2015}
Fortin, B. and Yazbeck, M. (2015).
\newblock Peer effects, fast food consumption and adolescent weight gain.
\newblock {\em Journal of Health Economics}, 42:125--138.

\bibitem[Friggeri et~al., 2014]{friggeri_rumor_2014}
Friggeri, A., Adamic, L., Eckles, D., and Cheng, J. (2014).
\newblock Rumor cascades.
\newblock {\em Proceedings of the 8th International AAAI Conference on Weblogs
  and Social Media (ICWSM)}.

\bibitem[Glazerman et~al., 2003]{glazerman_nonexperimental_2003}
Glazerman, S., Levy, D.~M., and Myers, D. (2003).
\newblock Nonexperimental versus experimental estimates of earnings impacts.
\newblock {\em The Annals of the American Academy of Political and Social
  Science}, 589(1):63--93.

\bibitem[Goel et~al., 2015]{goel_structural_2015}
Goel, S., Anderson, A., Hofman, J., and Watts, D.~J. (2015).
\newblock The structural virality of online diffusion.
\newblock {\em Management Science}, 62(1):180--196.

\bibitem[Goto et~al., 2011]{goto_incentives_2011}
Goto, J., Aida, T., Aoyagi, K., and Sawada, Y. (2011).
\newblock Incentives and social preferences in a traditional labor contract:
  Evidence from rice planting experiments in the {P}hilippines.
\newblock {\em Journal of Behavioral Economics and Finance}, 4:94--96.

\bibitem[Granovetter, 1978]{granovetter_threshold_1978}
Granovetter, M. (1978).
\newblock Threshold models of collective behavior.
\newblock {\em American Journal of Sociology}, 83(6):1420--1443.

\bibitem[Guryan et~al., 2009]{guryan_peer_2009}
Guryan, J., Kroft, K., and Notowidigdo, M.~J. (2009).
\newblock Peer effects in the workplace: Evidence from random groupings in
  professional golf tournaments.
\newblock {\em American Economic Journal: Applied Economics}, 1(4):34--68.

\bibitem[Hahn, 1998]{hahn_role_1998}
Hahn, J. (1998).
\newblock On the role of the propensity score in efficient semiparametric
  estimation of average treatment effects.
\newblock {\em Econometrica}, 66(2):315--331.

\bibitem[Hastie et~al., 2008]{hastie_elements_2008}
Hastie, T., Tibshirani, R., and Friedman, J. (2008).
\newblock {\em The Elements of Statistical Learning: Data Mining, Inference,
  and Prediction}.
\newblock Springer, 2 edition.

\bibitem[Heckman et~al., 1997]{heckman_matching_1997}
Heckman, J.~J., Ichimura, H., and Todd, P.~E. (1997).
\newblock Matching as an econometric evaluation estimator: Evidence from
  evaluating a job training programme.
\newblock {\em The Review of Economic Studies}, 64(4):605 --654.

\bibitem[Heinsman and Shadish, 1996]{heinsman_assignment_1996}
Heinsman, D.~T. and Shadish, W.~R. (1996).
\newblock Assignment methods in experimentation: When do nonrandomized
  experiments approximate answers from randomized experiments?
\newblock {\em Psychological Methods}, 1(2):154--169.

\bibitem[Hemkens et~al., 2016]{hemkens_agreement_2016}
Hemkens, L.~G., Contopoulos-Ioannidis, D.~G., and Ioannidis, J.~P. (2016).
\newblock Agreement of treatment effects for mortality from routinely collected
  data and subsequent randomized trials: Meta-epidemiological survey.
\newblock {\em BMJ}, 352:i493.

\bibitem[Herbst and Mas, 2015]{herbst_peer_2015}
Herbst, D. and Mas, A. (2015).
\newblock Peer effects on worker output in the laboratory generalize to the
  field.
\newblock {\em Science}, 350(6260):545--549.

\bibitem[Hill, 2008]{hill_comment_2008}
Hill, J. (2008).
\newblock Comment.
\newblock {\em Journal of the American Statistical Association},
  103(484):1346--1350.

\bibitem[Hill et~al., 2004]{hill_comparison_2004}
Hill, J.~L., Reiter, J.~P., and Zanutto, E.~L. (2004).
\newblock A comparison of experimental and observational data analyses.
\newblock {\em Applied Bayesian modeling and causal inference from
  incomplete-data perspectives: An essential journey with Donald Rubin's
  statistical family}, pages 49--60.

\bibitem[Hirano et~al., 2003]{hirano_efficient_2003}
Hirano, K., Imbens, G.~W., and Ridder, G. (2003).
\newblock Efficient estimation of average treatment effects using the estimated
  propensity score.
\newblock {\em Econometrica}, 71(4):1161--1189.

\bibitem[Imai and van Dyk, 2004]{imai_causal_2004}
Imai, K. and van Dyk, D.~A. (2004).
\newblock Causal inference with general treatment regimes: Generalizing the
  propensity score.
\newblock {\em Journal of the American Statistical Association},
  99(467):854--866.

\bibitem[Imbens, 2004]{imbens_nonparametric_2004}
Imbens, G.~W. (2004).
\newblock Nonparametric estimation of average treatment effects under
  exogeneity: A review.
\newblock {\em Review of Economics and Statistics}, 86(1):4--29.

\bibitem[Iyengar et~al., 2011]{iyengar_opinion_2011}
Iyengar, R., Van~den Bulte, C., and Valente, T.~W. (2011).
\newblock Opinion leadership and social contagion in new product diffusion.
\newblock {\em Marketing Science}, 30(2):195--212.

\bibitem[Katz and Lazarsfeld, 1955]{katz_personal_1955}
Katz, E. and Lazarsfeld, P.~F. (1955).
\newblock {\em Personal Influence: The Part Played by People in the Flow of
  Mass Communications}.
\newblock Free Press, New York.

\bibitem[King and Nielsen, 2015]{king_propensity_2016}
King, G. and Nielsen, R. (2015).
\newblock Why propensity scores should not be used for matching.
\newblock Working paper.

\bibitem[{LaLonde}, 1986]{lalonde_evaluating_1986}
{LaLonde}, R.~J. (1986).
\newblock Evaluating the econometric evaluations of training programs with
  experimental data.
\newblock {\em The American Economic Review}, 76(4):604--620.

\bibitem[Lewis and Rao, 2015]{lewis_unfavorable_2015}
Lewis, R.~A. and Rao, J.~M. (2015).
\newblock The unfavorable economics of measuring the returns to advertising.
\newblock {\em The Quarterly Journal of Economics}, 130(4):1941--1973.

\bibitem[Long et~al., 2008]{long_causal_2008}
Long, Q., Little, R.~J., and Lin, X. (2008).
\newblock Causal inference in hybrid intervention trials involving treatment
  choice.
\newblock {\em Journal of the American Statistical Association},
  103(482):474--484.

\bibitem[Lyle, 2007]{lyle_estimating_2007}
Lyle, D.~S. (2007).
\newblock Estimating and interpreting peer and role model effects from randomly
  assigned social groups at {West Point}.
\newblock {\em The Review of Economics and Statistics}, 89(2):289--299.

\bibitem[Manski, 1993]{manski_reflection_1993}
Manski, C.~F. (1993).
\newblock Identification of endogenous social effects: The reflection problem.
\newblock {\em The Review of Economic Studies}, 60(3):531 --542.

\bibitem[Manski, 2000]{manski_economic_2000}
Manski, C.~F. (2000).
\newblock Economic analysis of social interactions.
\newblock {\em The Journal of Economic Perspectives}, 14(3):115--136.

\bibitem[{McPherson} et~al., 2001]{mcpherson_birds_2001}
{McPherson}, M., {Smith-Lovin}, L., and Cook, J.~M. (2001).
\newblock Birds of a feather: Homophily in social networks.
\newblock {\em Annual Review of Sociology}, 27:415--444.

\bibitem[Michalopoulos et~al., 2004]{michalopoulos_can_2004}
Michalopoulos, C., Bloom, H.~S., and Hill, C.~J. (2004).
\newblock Can propensity-score methods match the findings from a random
  assignment evaluation of mandatory welfare-to-work programs?
\newblock {\em Review of Economics and Statistics}, 86(1):156--179.

\bibitem[Montanari and Saberi, 2010]{montanari_spread_2010}
Montanari, A. and Saberi, A. (2010).
\newblock The spread of innovations in social networks.
\newblock {\em Proceedings of the National Academy of Sciences}, 107(47):20196
  --20201.

\bibitem[Morgan and Harding, 2006]{morgan_matching_2006}
Morgan, S.~L. and Harding, D.~J. (2006).
\newblock Matching estimators of causal effects: Prospects and pitfalls in
  theory and practice.
\newblock {\em Sociological Methods \& Research}, 35(1):3--60.

\bibitem[Myers et~al., 2012]{myers_information_2012}
Myers, S.~A., Zhu, C., and Leskovec, J. (2012).
\newblock Information diffusion and external influence in networks.
\newblock In {\em Proceedings of the 18th ACM SIGKDD international conference
  on Knowledge discovery and data mining}, pages 33--41. ACM.

\bibitem[Owen and Eckles, 2012]{owen_bootstrapping_2012}
Owen, A.~B. and Eckles, D. (2012).
\newblock Bootstrapping data arrays of arbitrary order.
\newblock {\em The Annals of Applied Statistics}, 6(3):895--927.

\bibitem[Pohl et~al., 2009]{pohl_unbiased_2009}
Pohl, S., Steiner, P.~M., Eisermann, J., Soellner, R., and Cook, T.~D. (2009).
\newblock Unbiased causal inference from an observational study: Results of a
  within-study comparison.
\newblock {\em Educational Evaluation and Policy Analysis}, 31(4):463--479.

\bibitem[Robins and Rotnitzky, 1995]{robins_semiparametric_1995}
Robins, J.~M. and Rotnitzky, A. (1995).
\newblock Semiparametric efficiency in multivariate regression models with
  missing data.
\newblock {\em Journal of the American Statistical Association},
  90(429):122--129.

\bibitem[Rosenbaum and Rubin, 1983]{rosenbaum_central_1983}
Rosenbaum, P. and Rubin, D.~B. (1983).
\newblock The central role of the propensity score in observational studies for
  causal effects.
\newblock {\em Biometrika}, 70(1):41 --55.

\bibitem[Rosenbaum and Rubin, 1984]{rosenbaum_reducing_1984}
Rosenbaum, P.~R. and Rubin, D.~B. (1984).
\newblock Reducing bias in observational studies using subclassification on the
  propensity score.
\newblock {\em Journal of the American Statistical Association},
  79(387):516--524.

\bibitem[Rubin, 1997]{rubin_estimating_1997}
Rubin, D.~B. (1997).
\newblock Estimating causal effects from large data sets using propensity
  scores.
\newblock {\em Annals of Internal Medicine}, 127(8 Part 2):757--763.

\bibitem[Sacerdote, 2001]{sacerdote_peer_2001}
Sacerdote, B. (2001).
\newblock Peer effects with random assignment: Results for {D}artmouth
  roommates.
\newblock {\em Quarterly Journal of Economics}, 116(2):681--704.

\bibitem[Salganik et~al., 2006]{salganik_experimental_2006}
Salganik, M.~J., Dodds, P.~S., and Watts, D.~J. (2006).
\newblock Experimental study of inequality and unpredictability in an
  artificial cultural market.
\newblock {\em Science}, 311(5762):854--856.

\bibitem[Schuemie et~al., 2014]{schuemie_interpreting_2014}
Schuemie, M.~J., Ryan, P.~B., DuMouchel, W., Suchard, M.~A., and Madigan, D.
  (2014).
\newblock Interpreting observational studies: Why empirical calibration is
  needed to correct p-values.
\newblock {\em Statistics in Medicine}, 33(2):209--218.

\bibitem[Shadish et~al., 2008]{shadish_can_2008}
Shadish, W.~R., Clark, H.~H., and Steiner, P.~M. (2008).
\newblock Can nonrandomized experiments yield accurate answers? {A} randomized
  experiment comparing random and nonrandom assignments.
\newblock {\em Journal of the American Statistical Association},
  103(484):1334--1344.

\bibitem[Shalizi and Thomas, 2011]{shalizi_homophily_2011}
Shalizi, C.~R. and Thomas, A.~C. (2011).
\newblock Homophily and contagion are generically confounded in observational
  social network studies.
\newblock {\em Sociological Methods \& Research}, 40(2):211--239.

\bibitem[Sherif, 1936]{sherif_psychology_1936}
Sherif, M. (1936).
\newblock {\em The Psychology of Social Norms}.
\newblock Harper, New York.

\bibitem[Smith and Todd, 2005]{smith_does_2005}
Smith, J.~A. and Todd, P.~E. (2005).
\newblock Does matching overcome {LaLonde}'s critique of nonexperimental
  estimators?
\newblock {\em Journal of Econometrics}, 125(1):305--353.

\bibitem[Steiner et~al., 2015]{steiner_bias_2015}
Steiner, P.~M., Cook, T.~D., Li, W., and Clark, M. (2015).
\newblock Bias reduction in quasi-experiments with little selection theory but
  many covariates.
\newblock {\em Journal of Research on Educational Effectiveness},
  8(4):552--576.

\bibitem[Steiner et~al., 2010]{steiner_importance_2010}
Steiner, P.~M., Cook, T.~D., Shadish, W.~R., and Clark, M.~H. (2010).
\newblock The importance of covariate selection in controlling for selection
  bias in observational studies.
\newblock {\em Psychological Methods}, 15(3):250.

\bibitem[Thomas, 2013]{thomas_social_2013}
Thomas, A.~C. (2013).
\newblock The social contagion hypothesis: comment on `{S}ocial contagion
  theory: {E}xamining dynamic social networks and human behavior'.
\newblock {\em Statistics in Medicine}, 32(4):581--590.

\bibitem[Ugander et~al., 2012]{ugander_structural_2012}
Ugander, J., Backstrom, L., Marlow, C., and Kleinberg, J. (2012).
\newblock Structural diversity in social contagion.
\newblock {\em Proceedings of the National Academy of Sciences},
  109(16):5962--5966.

\bibitem[van~de Waal et~al., 2013]{van_potent_2013}
van~de Waal, E., Borgeaud, C., and Whiten, A. (2013).
\newblock Potent social learning and conformity shape a wild primate's foraging
  decisions.
\newblock {\em Science}, 340(6131):483--485.

\bibitem[{VanderWeele}, 2011]{vanderweele_sensitivity_2011}
{VanderWeele}, T.~J. (2011).
\newblock Sensitivity analysis for contagion effects in social networks.
\newblock {\em Sociological Methods \& Research}, 40(2):240--255.

\bibitem[Varian, 2016]{varian_causal_2016}
Varian, H.~R. (2016).
\newblock Causal inference in economics and marketing.
\newblock {\em Proceedings of the National Academy of Sciences},
  113(27):7310--7315.

\bibitem[Whiten et~al., 2005]{whiten_conformity_2005}
Whiten, A., Horner, V., and De~Waal, F.~B. (2005).
\newblock Conformity to cultural norms of tool use in chimpanzees.
\newblock {\em Nature}, 437(7059):737--740.

\bibitem[Wong et~al., 2016]{wong_empirical_2016}
Wong, V., Valentine, J.~C., and Miller-Bains, K. (2016).
\newblock Empirical performance of covariates in education observational
  studies.
\newblock {\em Journal of Research on Educational Effectiveness},
  (just-accepted):00--00.

\bibitem[Wu et~al., 2011]{wu_who_2011}
Wu, S., Hofman, J.~M., Mason, W.~A., and Watts, D.~J. (2011).
\newblock Who says what to whom on twitter.
\newblock In {\em Proceedings of the 20th international conference on World
  wide web}, {WWW} '11, pages 705--714. {ACM}.

\bibitem[Zimmerman, 2003]{zimmerman_peer_2003}
Zimmerman, D.~J. (2003).
\newblock Peer effects in academic outcomes: Evidence from a natural
  experiment.
\newblock {\em The Review of Economics and Statistics}, 85(1):9--23.

\end{thebibliography}

\clearpage

\end{document}